\documentclass[12pt]{iopart} 
 \usepackage{amssymb}
\usepackage{graphicx}
\usepackage{lipsum}
\usepackage{iopams}

\expandafter\let\csname equation*\endcsname\relax

\expandafter\let\csname endequation*\endcsname\relax

\usepackage{amsmath}
\usepackage{bbold}
\usepackage{bm}

\newcommand{\HH}{\mathcal{H}}

\begin{document}

       \title[Quantum properties of light scattered from many-body phases of ultracold atoms ]{Quantum properties of light scattered from structured many-body phases of ultracold atoms in quantum optical lattices}
         \author{Santiago F.  Caballero-Benitez$^1$, and Igor B. Mekhov}
         \ead{$^1$santiago.caballerobenitez@physics.ox.ac.uk}
           \address{University of Oxford, Department of Physics, Clarendon Laboratory, Parks Road, Oxford OX1 3PU, UK}
              \begin{abstract}
Quantum trapping potentials for ultracold gases change the landscape of classical properties of scattered light and matter.  The atoms in a quantum many-body correlated phase of matter change the properties of light and vice versa. The properties of both light and matter can be tuned by design and depend on the interplay between long-range (nonlocal) interactions  mediated by an optical cavity and short-range processes of the atoms.  Moreover, the quantum properties of light get significantly altered by this interplay, leading the light to have nonclassical features. Further, these nonclassical features can be designed and optimised.
\end{abstract}

       \maketitle

\section{Introduction.}
Optical lattices (OL's) offer the ultimate control of atoms trapped  by them. This leads to the formation of correlated phases of matter~\cite{Lewenstein}, this being useful for quantum simulation purposes~\cite{Bloch} and quantum information processing (QIP) applications. The degree of precision achieved with them so far, has allowed to achieved self-consistent light-matter states in a Bose-Einstein condensate (BEC) inside an optical cavity~\cite{EsslingerNat2010,HemmerichScience2012,ZimmermannPRL2014}.  Using the dynamical properties of the light~\cite{RitschRMP} the structural Dicke phase transition was achieved forming a state with supersolid features~\cite{EsslingerNat2010}.    However, the study of the full quantum regime of the system has been limited to few atoms~\cite{EPJD08,VukicsNJP2007,RitschLight,KramerRitschPRA2014, RitschArxiv2015}. As the light matter coupling is strongly enhanced in a high finesse optical cavity in a preferred wavelength, the atoms re-emit light comparable with the lasers used in the trapping process. As a consequence, an effective long-range (nonlocal) interaction emerges driven by the cavity field. It is now experimentally possible to access the regime where light-matter coupling is strong enough and the cavity parameters allow  to study the formation of quantum many-body phases with cavity decay rates of MHz~\cite{PNASEsslinger2013} and kHz~\cite{PNASHemmerich2015}. The light inside the cavity can be used to control the formation of many-body phases of matter even in a single cavity mode~\cite{EPJD08,MekhovNP2007,MekhovRev,Atoms}. This leads to several effects yet to be observed due to the dynamical properties of light~\cite{Larson,Morigi,Hofstetter,Reza,Morigi2}. Moreover, it has been shown that multimode atomic density patterns can emerge, even their coherences can become structured and light-matter quantum correlations can control the formation of correlated phases. Thus, a plethora of novel quantum phases due to the imprinting of structure by design in the effective light-induced interaction occurs~\cite{Santiago}. In addition to light-scattering~\cite{MekhovNP2007,PRLMekhov09}, homogenous quantum many-body phases can be measured by matter wave scattering~\cite{HPM1,HPM2,HPM3,HPM4} and dynamical structure factors can be obtained via homodyne detection~\cite{structureFMW}.
Recently, density ordering has been achieved with classical atoms~\cite{LabeyrieNaturePhotonics2014}. Further, multimode cavities extend the range of quantum phases even further~\cite{KramerRitschPRA2014,LevNaturePhys, Kollar2015, Muller2012}. Therefore, by carefully tuning system parameters and the spatial structure of light, one can design with plenty of freedom the quantum many-body phases that emerge. The quantum nature of the potential seen by the atoms changes the landscape of correlated quantum many-body phases  beyond  classical optical lattice setups. Very recently an optical lattice in an optical cavity has been realised~\cite{Esslinger2015} and self-oganized Mott-insultator phases have been achieved~\cite{Hemmerich2015} 

Moreover, the interplay between short range processes, such as on-site interactions and tunneling,  and  long-range cavity induced interactions can change significantly the properties of the light in the system. As these processes compete to optimise the energy in the system, the back-action of the matter affects the light generating nonclassical features~\cite{Knight,MorigiLight}. We show how such nonclassical effects of the light inside the cavity arise due to the emergence of structured quantum phases of matter. This can be traced back to the particular structure of the full light-matter state, which we construct beyond the limit where the light  can be integrated out (adiabatically eliminated)~\cite{EPJD08, Morigi,PRA2009}.
The formulation of the explicit form of the light-matter state of the system and deriving the effective matter Hamiltonian incorporating the effect of light at the quantum level is a difficult problem. 
We provide an alternative to those methods that allows for the construction of the effective Hamiltonian, where the effect of local processes (regular atomic tunneling and on site  interaction) is both considered in the properties of the light. The series of terms that arise, compose different hierarchies of light-induced interactions in addition to the adiabatic limit. 
This leads to a new effective Hamiltonian where the effect of local processes, such as tunneling and on-site interactions, and the global structure imprinted by the light is relevant. We use the technique of canonical transformations constructing the set of unitary operators to remove the non-diagonal terms due to the light~\cite{Mahan,Wagner}. The underlying symmetries broken by design by pumping light into the system modify the structure of both matter and light and the competition between global and local processes are the origin of nonclassical features. We find the effective matter Hamiltonian and that the full light-matter state is a superposition of squeezed coherent states. These depend on the emergent quantum many-body phases of matter the system supports and their structural properties. We demonstrate how the quantum  (quadratures) and classical (amplitude) properties of the light encode information about the strongly correlated phases of matter. As a corollary of our results, we find the conditions to optimise quadrature light squeezing in the system and the effect of the structure induced to the matter. Thus, our work will foster the design of this kind of states and their possible application towards quantum multimode systems in the analogous interdisciplinary field of optomechanics~\cite{RMP2014Optomech}.  Towards possible applications, there is an active interest in achieving large light squeezing in optomechanical systems where relevant achievements have already been made~\cite{PRXRegal}.  Recently, using trapped ions~\cite{IonScience} superpositions of squeezed states have been achieved as proposed by~\cite{CiracSqueezed}. Additionally, stationary entanglement of photons and atoms in a cavity has been studied~\cite{Morigimode}, seeding patters via the cavity field~\cite{MorigiRitsch}  and quantum control projection~\cite{BuchleitnerControl,ShersonControl}, opening the venue for applications on QIP. Beyond the quantum properties of light and matter, we find the effective master equation that describes the evolution of the system. This enables the possibility to study the effect of measurement back-action and its direct interplay with local processes.   Additionally, this can be used for state preparation using state projection via measurement  back-action~\cite{LP10,LP11,Gabriel}, while engineering of non-trivial correlated quantum states is possible~\cite{GabrielAFM,WojciechNHQZ}, and opens the possibility to optimise nonclassical properties of light.
\begin{figure}[htbp!]
\begin{center}
\includegraphics[width=0.8\textwidth]{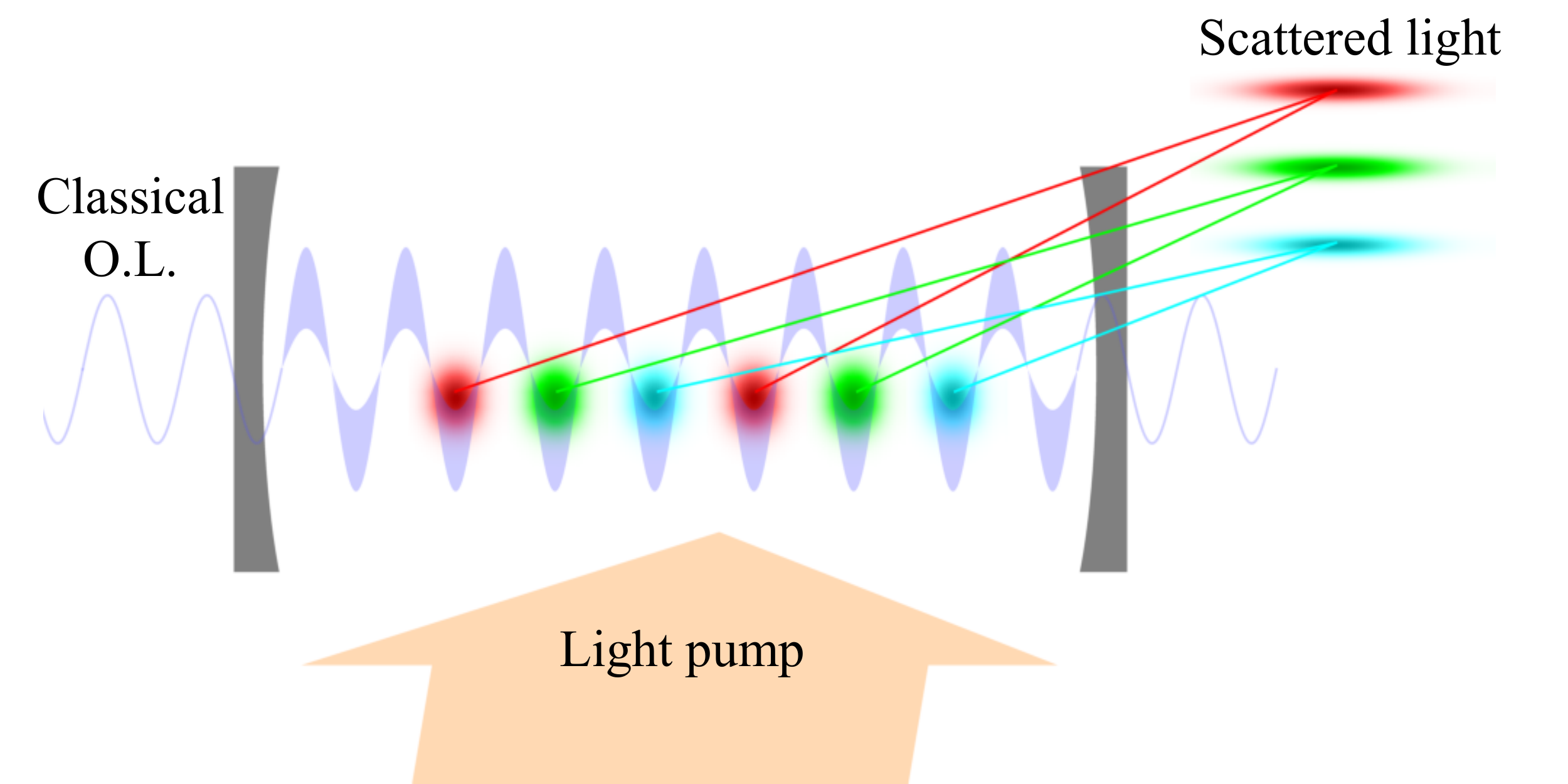}
\end{center}
\caption{{Cold atoms trapped in an optical lattice subject to a quantum potential created by the light inside a single-mode cavity.} The unsharp potential contour schematically depicts quantum fluctuations of light, which induce the light-matter correlations on top of the classical optical lattice potential, created by external laser beams. The cavity can be a standing- or traveling-wave. Different colours represent atoms corresponding to different light-induced spatially structured atomic modes. The superposition of squeezed coherent states corresponding to each light-induced mode is depicted on the right.}
\label{FQP}
\end{figure}
\section{The system.}
The system consists of atoms trapped in an OL inside single-mode cavity with the mode frequency $\omega_c$ and decay rate $\kappa$ in off-resonant scattering. The pump light  has amplitude  $\Omega_p$ (in units of the Rabi frequency) and frequency $\omega_p$ ($\Delta_c=\omega_p-\omega_c$). The system is illuminated in a plane transverse to the cavity axis (not necessarily at $90^\circ$).   The cavity mode couples with the atoms via the effective coupling strength $g_2= g \Omega_p/(2\Delta_a)$, with $g$  the light-matter coupling coefficient and $\Delta_a$ is the detuning between the light and atomic resonance~\cite{MekhovRev,Gabriel,Wojciech}. This can be described by the Hamiltonian $\HH=\HH^b+\HH^{a}+\HH^{ab}$, where $\HH^b$ is the regular Bose-Hubbard (BH) Hamiltonian~\cite{Fisher,Dieter},
\begin{equation}
\HH^b=-t_0\sum_{\langle i, j\rangle}(\hat b^\dagger_i\hat b^{\phantom{\dagger}}_j+h.c)-\mu\sum_i\hat n_i+\frac{U}{2}\sum_i\hat n_i(\hat n_i-1),
\end{equation}
with $t_0$ the nearest neighbour tunneling amplitude, $U$ the on-site interaction and $\mu$ the chemical potential.
The light is described by  $\HH^{a}=\hbar\omega_c\hat a^\dagger \hat a$ and the light-atom interaction is~\cite{MekhovRev}:
\begin{equation}
\HH^{ab}=g_2^*\hat a\hat F^\dagger+g_2\hat a^\dagger\hat F
\label{LMp}
\end{equation}
with
$
\hat F= \hat D+\hat B$. $\hat D=\sum_{j}J_{j,j}\hat n_j$ is the density coupling of light to the atoms, $\hat B=\sum_{\langle i,j\rangle}J_{i,j}( \hat b^\dagger_i\hat b^{\phantom{\dagger}}_j+h.c.)$ is due to the inter-site densities reflecting matter-field interference, or bonds~\cite{Santiago,Wojciech}. The sums go over illuminated sites $N_s$, and nearest neighbour pairs $\langle i,j\rangle$. The operators $b_i^\dagger$ ($\hat b_i$) create (annihilate) bosonic atoms at site $i$, $\hat a^\dagger$  ($\hat a$)  photons in the cavity, while the number operator of atoms per site is given by $\hat n_i=\hat b_i^\dagger\hat b_i^{\phantom{\dagger}}$. $\HH^{ab}$  is the relevant contribution to the quantum potential seen by atoms on top of classical OL described by the BH model, where the on-site interaction $U$ and hopping amplitude $t_0$ are short-range local processes.
{  The effective parameters of the Bose-Hubbard Hamiltonian with the cavity field can be calculated from the Wannier functions and are given by
\begin{eqnarray}
t_0&=&\int w(\mathbf{x}-\mathbf{x}_i)(\nabla^2-V_{OL}(\mathbf{x}))w(\mathbf{x}-\mathbf{x}_j)\mathrm{d}^n x,
\nonumber\\
\\
J_{i,j}&=&\int w(\mathbf{x}-\mathbf{x}_i)u^*_c(\mathbf{x})u_p(\mathbf{x})w(\mathbf{x}-\mathbf{x}_j)\mathrm{d}^n x,
\end{eqnarray}
where $u_{c,p}(\mathbf{x})$ are the cavity and pump mode functions and $w(\mathbf{x})$ are the Wannier functions. The classical optical lattice potential is given by $V_{OL}(\mathbf{x})$. }
 The classical optical lattice defining the regular Bose-Hubbard Hamiltonian is weakly dependent of the cavity parameters. The atoms are mainly trapped by the strong classical lattice, which is created inside a cavity by external laser beams. This external potential is insensitive to the quantum state of atoms. The light scattered into the cavity constitutes a quantum perturbation of the strong classical potential. This perturbation strongly depends on the many-body atomic state. Additional nonlinear dependence on the Bose Hubbard parameters leading to semi-classical effects can be incorporated as in~\cite{Larson,Morigi}. Further, the classical optical lattice and cavity light can be detuned from each other.  The light is pumped from the side of the main axis of the high Q cavity, at an angle not necessarily at $90^\circ$. The system is depicted in Fig.\ref{FQP}, where the effect on the scattered light is shown and will be explained through the paper. 

Moreover, it is useful to exploit the spatial structure of light as a natural basis to define atomic modes, as the coupling coefficients $J_{i,j}$ can periodically repeat in space~\cite{Santiago, Gabriel, Thomas,Wojciech}.  All atoms equally coupled to light belong to the same mode, while the ones coupled differently belong to different modes $\varphi$. Then we have for the atomic operators,
\begin{equation}
\hat F=\sum_{\varphi}J_{D,\varphi}\hat N_\varphi+\sum_{\varphi'}J_{B,\varphi'}\hat S_{0,\varphi'}
\end{equation}
where the light induced ``density"  $\hat N_\varphi$ and  ``bond"  $\hat S_{0,\varphi}$ mode operators, such that:
\begin{equation}
\hat N_\varphi=\sum_{i\in\varphi}\hat n_i ,\; \textrm{and}\; \hat S_{0,\varphi}=\sum_{\langle i,j\rangle\in\varphi}(\hat b_i^\dagger\hat b_j^{\phantom{\dagger}}+\hat b_j^\dagger\hat b_i^{\phantom{\dagger}}),
\end{equation}
{ 
with $J_{D,\varphi}$ corresponding to the posible values of $J_{i,i}$ and $J_{B,\varphi'}$ corresponding to $J_{i,j}$ where the pair $\langle i,j\rangle$ are nearest neighbours. For example, when illuminating in the diffraction minima $J_{D,\varphi_\pm}=\pm J_D$ and $J_{B,\varphi_\pm}=\pm J_B$ where $J_{D/B}$ are some constants. These encompass the different sets of values taken by the Wannier overlap integrals $J_{i,j}$~\cite{Santiago, Gabriel, Thomas,Wojciech}. In addition, it is useful to define new emergent mode structures $J_{E,\varphi}$ corresponding to $J^E_{i,j}=J_{j,j}-J_{i,i}$, with $\langle i,j\rangle$ nearest neighbours. In this case in diffraction minima, $J_{E,\varphi_\pm}=\pm2J_D$. Note that these emergent bond terms are absent for structureless light, that is, scattering as it happens in the diffraction maxima of light ($J_{i,i}=\textrm{const}$).
}

\section{Method}
\subsection{Hilbert space rotations.}
In order to describe the physics of the system, one  can construct an effective matter Hamiltonian from $\HH$. However, difficulty arises because the operators $\hat F$ and the BH Hamiltonian do not commute in general. Eliminating the light in the adiabatic limit, one can construct the effective Hamiltonian by different methods~\cite{EPJD08, Morigi,PRA2009}. However, the description is only accurate as long as the magnitude of the detuning $\Delta_c$  is very large compared with any other energy scale.
Beyond the adiabatic limit, our method relies on a series of canonical transformations constructed to eliminate the non-diagonal terms from the light-matter Hamiltonian. In our method, the cavity decay rate $\kappa$ has been introduced phenomenologically to reproduce the limit of adiabatic elimination of the light field. We find the additional corrections due to the non-commutativity the light-matter interaction with the local processes of the matter part of the Hamiltonian. Essentially, we perform a sequence of rotations on the Hilbert space using the formula,
\begin{equation}
\tilde\HH=\exp(-\hat R)\HH\exp(\hat R)=\HH+\sum_{n=1}^\infty\frac{\big[\HH,\hat R\big]_n}{n!}
\label{formula}
\end{equation}
with $\big[\HH,\hat R\big]_n$ the $n$-th order commutator with respect to $\hat R$. A sequence of rotation operators $\hat R$ is performed, where these are chosen to remove the non-diagonal part of the light field in the light-matter interaction via the commutator expansion after their action. The particular structures needed and the number of rotations depend on the underlying structure of the matter part Hamiltonian in the light-matter interaction and the BH model. Their interplay  with the by-products of each rotation determines the consecutive rotation operator to be constructed. Thus, one provides an ansatz for each $\hat R$ operator and eliminates according to the by-product of the next transformation~\cite{Wagner}.  The result of our method is a theory that incorporates in a perturbative operator expansion the interplay of the non-commutative character between local processes (tunnelling and on-site interactions) and the long-range (nonlocal) light induced effective interactions. 

\subsubsection{Adiabatic limit rotation.}
The first transformation recovers the limit when light can be adiabatically eliminated, explicitly this is: 
\begin{equation}
\hat R_{\mathrm{ad}}=c\hat F\hat a^\dagger-c^*\hat F^\dagger\hat a.
\end{equation}
Therefore,
\begin{equation}
e^{\hat R_{\mathrm{ad}}}=\exp(c\hat F\hat a^\dagger-c^*\hat F^\dagger\hat a)=D(c\hat F )
\end{equation}
is analogous to the displacement operator from quantum optics~\cite{Knight, Walls, Noise}, $c=g_2/(\Delta_c+i\kappa)$ is the cavity Purcell factor. 
Thus using (\ref{formula}),
{  
\begin{eqnarray}
\HH'&=&e^{\hat R_{\mathrm{ad}}}\HH e^{-\hat R_{\mathrm{ad}}}
\\
&=&-\Delta_c\hat a^\dagger\hat a+\HH^b+\frac{g_{\mathrm{eff}}}{2}(\hat F^\dagger \hat F+\hat F\hat F^\dagger)+\frac{i g_{\mathrm{eff}}\kappa}{\Delta_c}\big[\hat F,\hat F^\dagger]
\label{adialim}
\\
&+&\big[\HH^b,\hat F\big]c\hat a^\dagger-\big[\HH^b,\hat F^\dagger\big]c^*\hat a
\label{dyneff}
\\
&+&\frac{1}{2}\big[\big[\HH^b,\hat F\big]c\hat a^\dagger-\big[\HH^b,\hat F^\dagger\big]c^*\hat a,\hat F\big]c\hat a^\dagger
\label{sq1}
\\
&-&\frac{1}{2}\big[\big[\HH^b,\hat F\big]c\hat a^\dagger-\big[\HH^b,\hat F^\dagger\big]c^*\hat a,\hat F^\dagger\big]c^*\hat a+\dots
\label{sq2}
\end{eqnarray}
to remove the additional non-diagonal light terms beyond the adiabatic limit (\ref{adialim}) we need to perform additional rotations as we state below. Dynamical effects occur due to (\ref{dyneff}) and light squeezing originates from (\ref{sq1}) and (\ref{sq2}), while the dots refer to higher order photon processes due to the commutator expansion that will not be considered in what follows.}
\subsubsection{Rotations due to dynamical effects.} 
A subsequent series of rotations is employed to eliminate term by term the emergence of non diagonal terms in the light sector due to the interplay with short range processes. The sequence of rotations is,
\begin{equation}
\HH''=e^{\hat R_{\chi,\infty}}\cdots e^{\hat R_{\chi,1}}e^{\hat R_{\chi,0}}\HH'e^{-\hat R_{\chi,0}}e^{-\hat R_{\chi,1}}\cdots e^{-\hat R_{\chi,\infty}}
\end{equation}
where each rotation is given by,
\begin{equation}
\hat R_{\chi,k}=\tilde{U}^k(c\hat\chi_k\hat a^\dagger+c^*\hat\chi_k^\dagger\hat a),
\end{equation}
 with $k\in\mathbb{Z}^+_0$, where, 
 $\hat\chi_n=\tilde{U}\hat g_n+\tilde{t}_0\hat f_n$, such that 
 \begin{equation}
 \hat g_n=\sum_{k=0}^n\binom{n}{k} \hat z_k \textrm{ and } \hat f_n=\sum_{k=0}^n\binom{n}{k} \hat y_k.
 \end{equation} Additionally, 
we cast our results using the natural choice of dimensionless expansion parameters, {  which are defined as: $\tilde{t}_0=t_0/\Delta_c$ and $\tilde{U}=U/\Delta_c$}, while
\begin{eqnarray}
&&\hat z_k=\sum_\varphi J_{B,\varphi}\hat S_{k,\varphi}\;\textrm{and}\; \hat y_k=\sum_\varphi J_{E,\varphi}\hat J_{k,\varphi},\; k \textrm{ even},
\nonumber\\
&&\hat z_k =\sum_\varphi J_{B,\varphi}\hat J_{k,\varphi}\;\textrm{and}\;\hat y_k =\sum_\varphi J_{E,\varphi}\hat S_{k,\varphi},\;  k \textrm{ odd},
\end{eqnarray}
where we have used  collective weighted ``bond" operators $\hat S_{k,\varphi}$ and weighted ``bond current" operators $\hat J_{k,\varphi}$ corresponding to the light induced modes $\varphi$. These are defined as, 
\begin{eqnarray}
\hat J_{k,\varphi}=\sum_{\langle i,j\rangle\in\varphi}\Delta \hat n_{i,j}^k(\hat b^\dagger_i\hat b^{\phantom{\dagger}}_j-\hat b^\dagger_j\hat b^{\phantom{\dagger}}_i)
\\
\hat S_{k,\varphi}=\sum_{\langle i,j\rangle\in\varphi}\Delta \hat n_{i,j}^k(\hat b^\dagger_i\hat b^{\phantom{\dagger}}_j+\hat b^\dagger_j\hat b^{\phantom{\dagger}}_i)
\end{eqnarray}
with $\Delta \hat n_{i,j}=\hat n_j-\hat n_i$. Thus these collective operators are spatially modulated by the difference in density between nearest neighbour pairs $\langle i, j \rangle$. The above operators can be traced back to  the modulations to the densities and currents that arise due to the fact that short range tunneling and/or on-site interactions do not commute with light-matter interaction components. The particular binomial structure of $\hat g_n$ and $\hat f_n$ arises as each rotation operator needed to diagonalize the Hamiltonian generates higher order operator polynomials terms recursively due to the commutator expansion.  These operators correspond to the emergence of particle-hole excitations and matter self-interactions in analogy to the Feynman diagram expansion in momentum space. {   Considering $\tilde t_0$ and $\tilde U$ as expansion parameters it is enough to consider the first few terms in the expansion of order ($O(\cdot)$) linear in $\tilde t_0$, $\tilde U$ and their product. Thus $R_{\chi,0}$ and $R_{\chi,1}$ are the leading terms, however the full expansion can be used depending on the coupling strengths and partial re-summation of families of terms can be employed as in the standard techniques of many-body physics \cite{Mahan,Wagner}.    
}

\subsubsection{Squeezing rotation.}
The final transformation to obtain the effective matter Hamiltonian of the system is achieved by,
\begin{equation}
e^{\hat R_\xi}=\exp[(\hat\xi^*\hat a^2-\hat\xi\hat a^{\dagger 2})/2]=S(\hat \xi)
\end{equation}
which is the analogous squeezing operator from quantum optics~\cite{Knight, Walls, Noise}. The squeezing amplitude operator is given by,
\begin{equation}
\hat \xi=-c^2\tilde{t}_0\sum_{\varphi'}J_{E,\varphi'}^2\hat S_{0,\varphi'}
-\frac{c^2}{2}\tilde{U}\sum_\varphi J_{B,\varphi}^2(\Delta \hat{N}_\varphi+\Delta \hat{J}_\varphi),
\end{equation} 
where it is useful to define ``density fluctuations" operators $\Delta \hat N_\varphi$ and ``bond current fluctuations" $\Delta\hat J_\varphi$,
 \begin{equation}
\Delta \hat N_{\varphi}=\sum_{\langle i,j\rangle\in\varphi}\Delta \hat n_{i,j}^2
\;\textrm{and}\;
\Delta \hat J_{\varphi}=\sum_{\langle i,j\rangle\in\varphi}(\hat b^\dagger_i\hat b^{\phantom{\dagger}}_j-\hat b^\dagger_j\hat b^{\phantom{\dagger}}_i)^2,
\end{equation}
These operators of quantum fluctuations of matter arise due to higher order light processes entangling the matter and the light. The transformation is used to eliminated non-diagonal terms due to higher order photon processes. We have restricted the expansion to two photon  processes. Therefore this contains the first nontrivial correction. The effect of additional higher order photon processes could be included in principle performing additional rotations, but we will not pursue this in what follows. 

After straight forward algebraic procedure applying each transformation and computing the relevant commutators using (\ref{formula}), we get the effective atomic matter Hamiltonian:
{  
\begin{equation}
 \HH_{\mathrm{eff}}=e^{\hat R_\xi}e^{\hat R_{\chi,\infty}}\cdots e^{\hat R_{\chi,0}}e^{\hat R_\mathrm{Ad}}\HH\e^{-\hat R_\mathrm{Ad}}e^{-\hat R_{\chi,0}}\cdots e^{-\hat R_{\chi,\infty}}e^{-\hat R_\xi}-\tilde{\HH}_a
 \end{equation} 
where, $\tilde{\HH}_a=-\Delta_c\hat a^\dagger\hat a$, which is the light part after the rotation to the pump frame of reference~\cite{EPJD08}.}

\section{Effective Hamitonian and Full light-matter state.}
\subsection{Effective Hamiltonian.}
The effective atomic Hamiltonian after the rotations is 
\begin{equation}
\HH_{\mathrm{eff}}=\HH_{\mathrm{ad}}+\HH_\xi+\HH_\chi,
\end{equation}
with,
\begin{equation}
\HH_{\mathrm{ad}}=\HH^b+\frac{g_\mathrm{eff}}{2}(\hat F\hat F^\dagger+\hat F^\dagger\hat F)
\label{ad}
\end{equation}
the result in the adiabatic limit of light~\cite{Santiago} with $|\kappa/\Delta_c|\ll1$, $|U/\Delta_c|\ll1$, $|t_0/\Delta_c|\ll1$  and $g_\mathrm{eff}=\Delta_c|g_2|^2/(\Delta_c^2+\kappa^2)=\Delta_c|c|^2$. {   Here we have neglected the non-abelian shift in (\ref{adialim}), but in general it can be non-zero depending on the light mode functions for complex values i.e. travelling wave configurations  where via either pump or cavity modes light in-between sites can be focused}. As it has been shown~\cite{Santiago,Wojciech}, this leads to the formation of structures of density and bond modes  that can be nearly independent from each other. It is possible to generate in a single mode cavity  spatial multimode structures of $R$ density modes~\cite{Gabriel,Thomas} and $2R$ bond modes~\cite{Santiago}  by carefully choosing how the light is pumped into the system~\cite{Wojciech}. In the adiabatic limit (\ref{ad}), the structure of matter is controlled by the interplay between the BH processes, regular atomic tunneling and on-site interaction, and the light induced interaction proportional to $g_{\mathrm{eff}}$. The ground state of (\ref{ad}) will be achieved whenever atoms scatter light  maximally for $g_{\mathrm{eff}}<0$ or minimally $g_{\mathrm{eff}}>0$~\cite{Santiago}. As we will show below, the additional terms $\HH_\xi$ are related to light squeezing and $\HH_\chi$ arises due to the dynamical corrections from the light induced processes and their interplay with the short-range BH processes due to the structure imprinted on the matter. These will modify the landscape of quantum phases the system can access, as well as, the properties of light beyond being a superposition of structured coherent states~\cite{Santiago}. 
Using the light induced mode decomposition, then we can write,
\begin{eqnarray}
\hat F^\dagger \hat F+\hat F \hat F^\dagger&=&
\sum_{\varphi,\varphi'}
[\gamma_{\varphi,\varphi'}^{D,D}
\hat N_\varphi^{\phantom{*}}
\hat N_{\varphi'}^{\phantom{*}}
+\gamma_{\varphi,\varphi'}^{B,B}
\hat S_{0,\varphi}^{\phantom{*}}\hat S_{0,\varphi'}^{\phantom{*}}
\nonumber
\\
&+&\gamma_{\varphi,\varphi'}^{D,B}(
\hat N_\varphi^{\phantom{*}}\hat S_{0,\varphi'}^{\phantom{*}}
+
\hat S_{0,\varphi'}^{\phantom{*}}\hat N_{\varphi}^{\phantom{*}})],
\label{mdecomp}
\end{eqnarray}
with $\gamma^{\nu,\eta}_{\varphi,\varphi'}=(J_{\nu,{\varphi}}^* J^{\phantom{*}}_{\eta,{\varphi'}}+c.c.)$, where we have used the light induced ``density"  $\hat N_\varphi$ and  ``bond"  $\hat S_{0,\varphi}$ mode operators.

The additional contributions in $\HH_{\mathrm{eff}}$ are the first order corrections in $\tilde{t}_0$ and $\tilde U$ originated by the non-commutative nature between local processes and the global structure introduced due to the light induced modes.  The terms due to light induced dynamics are
\begin{eqnarray}
\HH_\chi&=& g_{\mathrm{eff}}\tilde{t}_0\sum_{\varphi,\varphi'}(J_{E,\varphi}^*J_{D,\varphi'}^{\phantom{*}}-J_{D,\varphi'}^*J_{E,\varphi}^{\phantom{*}})\hat N_{\varphi'}\hat J_{0,\varphi}
\nonumber
\\
&+&\frac{g_{\mathrm{eff}}}{2}\sum_{n=0}^{\infty}\tilde{U}^{2n}(\hat \chi_n^\dagger\hat \chi_n+\hat \chi_n\hat \chi_n^\dagger)
\end{eqnarray}
These dynamical terms can have a significant effect on the effective Hamiltonian renormalizing the effective light induced interaction terms. Additionally, they can aid the formation of structured ground-states due to their dependency on the atom number difference between nearest neighbour sites.  In general, these terms tend to induce structure in the atomic density  as $|\tilde U|$ increases due to light-matter quantum correlations $g_\mathrm{eff}>0$  and due to semiclassical effects for $g_\mathrm{eff}<0$~\cite{Santiago}. In principle, going beyond the perturbation character of the expansion could be handled via renomalization.  Close to a structural phase transition, where $\langle\Delta\hat  n_{i,j}\rangle\approx 0$, for example, from a structured ground-state (with  DW order) to a homogenous ground-state (a normal superfluid),  the leading behaviour shows an instability for $|\tilde U|<1$ as,
\begin{equation}
\sum_{n=0}^{\infty}\tilde{U}^{2n}(\hat \chi_n^\dagger\hat \chi_n+\hat \chi_n\hat \chi_n^\dagger)\sim\frac{\hat \chi_0^\dagger\hat \chi_0+\hat \chi_0\hat \chi_0^\dagger}{1-\tilde U^2}
\end{equation}
The formation of this instability means a structure ground state can be an energetically favourable solution depending on the coupling constants strengths and the competition with other processes, from the adiabatic limit and the BH model. This provides an amazing potential for manipulation with the purpose of quantum simulation, as one can select the inhibition or enhancement of the interplay with local processes. One can design this using the structure constants $J_{i,j}$  and $g_{\mathrm{eff}}$, thus controlling the light induced mode formation and changing the onsite interactions via Feshbach resonances or even via the classical optical lattice potential.  The expansion could be further manipulated by the use of diagrammatic tools, including well know partial resumations techniques (i.e. Feynman diagrams) but we will not pursue this here, as we are interested in the regime where $\tilde t_0$ and $\tilde U$ are perturbation parameters. The terms due to the interplay with short-range processes $\HH_\chi$ and $\HH_\xi$ contain the effect of higher order correlated processes, as they contain in addition to 2-point correlations, $n$-point correlations with $n\geqslant4$.

Beyond the adiabatic limit and dynamically generated processes, the additional terms in $\HH_{\mathrm{eff}}$ modify the energy due to light squeezing are:
\begin{eqnarray}
\HH_\xi&=& g_{\mathrm{eff}} \tilde{t}_0\sum_{\varphi'}|J_{E,\varphi'}|^2\hat S_{0,\varphi'}
+\frac{g_{\mathrm{eff}}\tilde {U}}{2}\sum_\varphi |J_{B,\varphi}|^2(\Delta \hat{N}_\varphi+\Delta \hat{J}_\varphi).
\end{eqnarray}
$\Delta \hat N_{\varphi}$ and $\Delta \hat J_{\varphi}$ are strongly smeared out in the limit where the effect of the light is classical. This occurs because the atoms maximise light scattering to reach the ground-state of the effective Hamiltonian. However, these are relevant for the case where the strong classical signal is suppressed and the effect of light-matter quantum correlations is significant~\cite{Santiago}. In general, they have a suppression effect upon fluctuations for $U>0$ while they promote an instability for $U<0$, as $g_{\mathrm{eff}}\tilde{U}=|c|^2U$. In addition, they introduce a direct modification to tunneling amplitudes that are controlled by the pattern of emergent bond modes via the different possible couplings $J_{E,\varphi}$. Importantly, as we will show their origin has a non-trivial effect on the properties of light.

\subsection{Full Light-Matter state.}
As shown previously,  the effective Hamiltonian $\HH_{\mathrm{eff}}$ is diagonal in the light sector and first order quantum corrections have been included. It follows that the full solution to the light-matter state can be written as:
\begin{equation}
|\Psi\rangle=\sum_{\varphi_q}\Gamma^b_{\varphi_q}(t)\beta_{\varphi_q}|\varphi_q\rangle_b|\alpha^{\phantom{\chi}}_{\varphi_q}+\alpha^\chi_{\varphi_q},\xi_{\varphi_q}\rangle_a,
\end{equation}
where the subscript ``$a$" (``$b$") corresponds to the light (matter) part; $\Gamma^b(t)=\exp(-i\HH_{\mathrm{eff}}t)$, and $\Gamma^b_{\varphi_q}|\varphi_q\rangle_b=\hat\Gamma^b|\varphi_q\rangle_b$. The light components are squeezed coherent states  $|\alpha,\xi\rangle_a=D(\alpha)S(\alpha)|0\rangle_a$, with the squeezing operator $S(\xi)=\exp[(\xi^*\hat a^2-\xi\hat a^{\dagger 2})/2]$ and the displacement operator $D(\alpha)=\exp(\alpha\hat a^\dagger-\alpha^*\hat a)$~\cite{Knight,Walls,Noise}.  The ground state of the effective Hamiltonian is $|\Psi\rangle_b=\sum_{\varphi_q}|\varphi_q\rangle_b$. The light amplitudes due to the projection of the matter structure are $\alpha_{\varphi_q}|\varphi_q\rangle_b=c\hat F|\varphi_q\rangle_b$, $\alpha^\chi_{\varphi_q}|\varphi_q\rangle_b=c\hat\eta|\varphi_q\rangle_b$, with 
$\hat \eta=\sum_{n=0}^{\infty}\tilde{U}^{n}\hat \chi_n$.
The weights due to the dynamical character of the light are 
$
\beta_{\varphi_q}=\exp(|c|^2\sum_{n=0}^{\infty}\tilde{U}^{2n}|\chi_{n,\varphi_q}|^2),
 $
 with $\chi_{n,\varphi_q}|\varphi_q\rangle_b=\hat\chi_n|\varphi_q\rangle_b$. 
In addition, the squeezing parameter amplitudes corresponding to the projection onto the matter sector are $\xi_{\varphi_q}|\varphi_q\rangle_b=\hat\xi|\varphi_q\rangle_b$ with,
the squeezing amplitude operator $\hat\xi$.  Therefore, the structure of the strongly correlated matter gets imprinted in the quantum properties of light  via the squeezing parameter projections $\xi_{\varphi_q}|\varphi_q\rangle_b$. This generates a non-trivial superposition of squeezed coherent states entangled with the strongly correlated matter. In the above, we have neglected next-nearest neighbour and  higher processes, considered all the first order corrections $O(\tilde t_0)$ and terms order $O(J_{B,\varphi}^2)$ while constraining to see the effect of two photon processes. However, this is not a limitation in our method, since additional $n$-neighbour processes or additional $n$-photon processes can be incorporated straightforward if relevant. 

\section{Quantum properties of the scattered light.} 
\subsection{Photon number.}
The number of photons can be written as,
 \begin{eqnarray}
 \langle\hat a^\dagger\hat a\rangle&=&\sum_{q=1}^{N_R}|c_{\varphi_q}|^2(\sinh(r_{\varphi_q})^2+|\tilde\alpha_{\varphi_q}|^2)
 \nonumber\\
 &=&\langle\sinh(|\hat\xi|)^2\rangle+\frac{g_{\mathrm{eff}}}{2\Delta_c}\langle\hat G^\dagger\hat G+\hat G\hat G^\dagger\rangle
 \end{eqnarray}
 where $\hat G=\hat F+\hat\eta$ and  $|c_{\varphi_q}|^2=\beta_{\varphi_q}^2|_b\langle \varphi_q|\Psi\rangle_b|^2$  the weights corresponding to the matter component projections (the probabilities).  $\tilde\alpha_{\varphi_q}=\alpha^{\phantom{\chi}}_{\varphi_q}+\alpha^\chi_{\varphi_q}$ are the coherent state components and $r_{\varphi_q}=|\xi_{\varphi_q}|$ the corresponding squeezing parameters depending on the projections on the matter states, and $N_R$ the number of light induced components. Therefore the quantum properties of the matter are accessible at the level of the photon number. Moreover the light-amplitude is sensitive to the particular structure that emerges due to the correlated phases of matter~\cite{Santiago}.  In the limit of large detunning $|\Delta_c|\gg\{t_0,U\}$ the above reduces to,
 \begin{equation}
 \langle\hat a^\dagger\hat a\rangle\approx\frac{g_{\mathrm{eff}}}{2\Delta_c}\langle\hat F^\dagger\hat F+\hat F\hat F^\dagger\rangle
 \end{equation} 
 which is equivalent to the adiabatic limit~\cite{Santiago}.
 \begin{figure}[tp!]
\begin{center}
\includegraphics[width=0.95\textwidth]{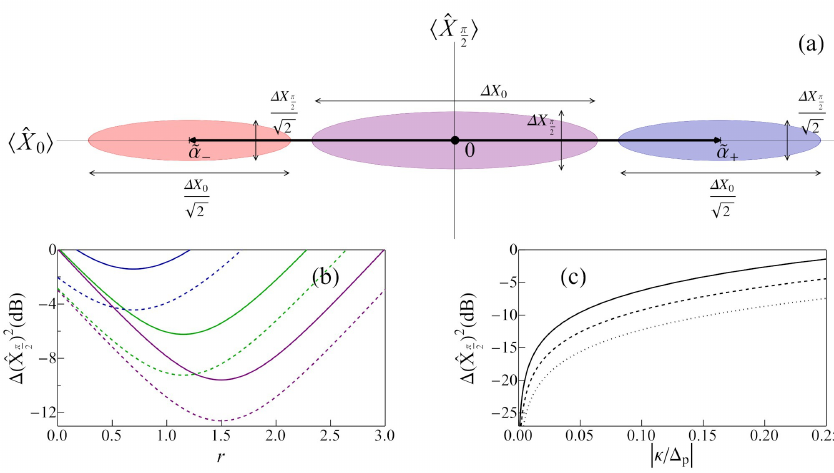}
\end{center}
\caption{(a) Quadrature components for light-induced two mode density coupling. Quadrature components are centred around $\tilde\alpha_\pm$ (red and blue), the quadrature is centred around zero (purple). The quadrature widths are given by $\Delta X_0\approx e^r(1+2[\tilde{\sigma}_p- \frac{\kappa^2}{\Delta_c^2}\sinh(r)\cosh(r)/(1+\frac{\kappa^2}{\Delta_c^2})]e^{-2r})^{1/2}/2$ and $\Delta X_{\frac{\pi}{2}}=e^{-r}(1+2\frac{\kappa^2}{\Delta_c^2}[\tilde{\sigma}_p+\sinh(r)\cosh(r)/(1+\frac{\kappa^2}{\Delta_c^2})]e^{2r})^{1/2}/2$, with $r$ the squeezing parameter, $\tilde\sigma_p=\sigma_p(\Delta(\hat n_+)^2+\Delta(\hat n_-)^2)$ and $\sigma_p=( g_{\mathrm{eff}} \Delta_c N_s J_D^2)/(\Delta_c^2+\kappa^2)$, with $|\kappa/\Delta_c|\ll1$. The system exhibits squeezing in the quadrature $\phi=\pi/2$. Each coherent state component is squeezed at the same angle. (b) Quadrature squeezing at $\phi=\pi/2$ in dB,  $|\kappa/\Delta_c|=0.25 \textrm{(blue)}, 0.1 \textrm{(green), 0.05 \textrm{(purple)}}$, dashed lines correspond to squeezing projecting to one component, solid lines to the full state. (c) Optimal Squeezing as a function of the cavity decay rate, optimal squeezing is achieved whenever $r=\ln(|{\Delta/\kappa}|)/2$ solid line. Dashes correspond to projecting to one component with two light induced modes, dots corresponds to projecting to a single component with 4 light induced modes. Parameters are: $\tilde\sigma_p=1$ (b) and (c).}
\label{Qpl}
\end{figure}
\subsection{Light Quadratures and Squeezing.}
 The squeezing in the quadratures can be written as, 
 \begin{eqnarray}
 \Delta(\hat X_\phi)^2&=&\frac{1}{4}\sum_{q=1}^{N_R}|c_{\varphi_q}|^2(1+2\sinh(r_{\varphi_q})^2)
 \nonumber\\
 &-&
  \frac{1}{2}\sum_{q=1}^{N_R}|c_{\varphi_q}|^2 \cos(\theta-2\phi)\cosh(r_\varphi)\sinh(r_\varphi)
\nonumber\\
& +&
 \frac{1}{4}\sum_{q=1}^{N_R}|c_{\varphi_q}|^2(e^{-i\phi}\tilde\alpha_{\varphi_q}+e^{i\phi}\tilde\alpha_{\varphi_q}^*)^2
 \nonumber\\
 &-&\frac{1}{4}\left(\sum_{q=1}^{N_R}|c_{\varphi_q}|^2(e^{-i\phi}\tilde\alpha_{\varphi_q}+e^{i\phi}\tilde\alpha_{\varphi_q}^*)\right)^2
 \end{eqnarray}
 with $\hat X_\phi=(e^{-i\phi}\hat a+e^{i\phi}\hat a^\dagger)/2$,  
 $\Delta(\hat X_\phi)^2=\langle \hat  X_\phi^2\rangle-\langle\hat X_\phi^{\phantom{2}}\rangle^2$,  and $\theta=\mathrm{arg}(2i\kappa\Delta_c+\kappa^2-\Delta_c^2)$.  This can be rewritten as,
 \begin{eqnarray}
 \Delta(\hat X_\phi)^2&=&\frac{1}{4}+\frac{1}{2}\langle\sinh(|\hat \xi|)^2\rangle 
 \nonumber\\
 &-&
  \frac{1}{2}\cos(\theta-2\phi)\langle\cosh(|\hat\xi|)\sinh(|\hat\xi|)\rangle
\nonumber\\
& +&
 \frac{ c^2e^{-i2\phi}}{4}\Delta(\hat G)^2+\frac{ c^{*2}e^{i2\phi}}{4}\Delta(\hat G^\dagger)^2
\nonumber \\
 &+&\frac{|c|^2}{4}(\langle\hat G^\dagger\hat G+\hat G\hat G^\dagger\rangle-2 \langle\hat G^\dagger\rangle\langle\hat G\rangle)
 \end{eqnarray} 
  In particular when $\tilde\alpha_\varphi=cG_\varphi^{\phantom{*}}$ and  $\tilde\alpha_\varphi^*=c^*G_\varphi$ so that $\hat G$ is Hermitian, then:
\begin{eqnarray}
 \Delta(\hat X_\phi)^2&=&\frac{1}{4}+\frac{1}{2}\langle\sinh(|\hat \xi|)^2\rangle 
 \nonumber\\
 &-&
  \frac{1}{2}\cos(\theta-2\phi)\langle\cosh(|\hat\xi|)\sinh(|\hat\xi|)\rangle
\nonumber\\
& +&
 \frac{g_{\mathrm{eff}}(\Delta_c\cos(\phi)-\kappa\sin(\phi))^2}{\Delta_c(\Delta_c^2+\kappa^2)}\Delta(\hat G)^2
 \end{eqnarray}
 For two light induced modes with density coupling ($J_{B,\varphi}=0$,$J_{D,\varphi}\neq0$) in mean-field approximation, the above reduces to: 
\begin{eqnarray}
 \Delta(\hat X_0)^2
 &\approx&
\frac{e^{2r}}{4}-\left(\frac{\kappa^2}{\Delta_c^2}\right)\frac{\sinh(r)\cosh(r)}{1+\frac{\kappa^2}{\Delta_c^2}}
\nonumber\\
&+&\frac{\sigma_p}{2}(\Delta(\hat n_+)^2+\Delta(\hat n_-)^2+O(\tilde t_0))
\\
 \Delta(\hat X_{\frac{\pi}{2}})^2
 &\approx&
\frac{e^{-2r}}{4}+\left(\frac{\kappa^2}{\Delta_c^2}\right)\frac{\sinh(r)\cosh(r)}{1+\frac{\kappa^2}{\Delta_c^2}}
\nonumber
\\
&+&
\frac{\sigma_p}{2}\left(\frac{\kappa^2}{\Delta_c^2}\right)(\Delta(\hat n_+)^2+\Delta(\hat n_-)^2+O(\tilde t_0)),
\nonumber
\\
\label{sqe}
 \end{eqnarray} 
 where $r=|\langle\hat\xi\rangle|$, and $\sigma_p=( g_{\mathrm{eff}} \Delta_c N_s J_D^2)/(\Delta_c^2+\kappa^2)$. The additional terms to regular squeezing are due to atomic fluctuations in each light induced component.  We have used the identities, 
 \begin{eqnarray}
& \frac{e^{\pm2 r}}{2}={\sinh (r)^2}\pm\sinh (r) \cosh (r)+\frac{1}{2},&
 \\
& \cos \left(\arg \left((y+i)^2\right)+
\frac{(1\pm1)\pi}{2}\right)=\pm\left(\frac{2 y^2}{y^2+1}-1\right),&
 \end{eqnarray}
 and that in mean-field approximation $\Delta(\hat N_+-\hat N_-)^2\approx N_s (\Delta(\hat n_-)^2+\Delta(\hat n_+)^2)$ \cite{Santiago}. 
 Beyond mean-field approximation, as the system goes in the normal SF state ($|t_0/U|\gg 0$) additional corrections due to additional coherent amplitudes that depend on their atomic fluctuations will increase the super-poissonian character of the light quadratures. The relation between quadrature components is shown in Fig.\ref{Qpl}a.  In particular, the quadratures at $\phi=\pi/2$ can be squeezed several dB in the case where the cavity detunning is of the order of the recoil energy $E_R$, as in~\cite{PNASHemmerich2015} with $|\kappa/\Delta_c|\approx 0.1-0.25$ and $|\Delta_c|\sim1-100E_R$. Choosing $|g_{\mathrm{eff}}|\sim E_R/N_s$  and $\sigma_p\sim 1$ with $\rho=3/2$ in the SF state where fluctuations are maximal, one can easily achieve $r\sim 1.5$ and about 10dB of squeezing improving the ratio $|\kappa/\Delta_c|\approx 0.05$ by changing the detunning or improving the cavity, see Fig.\ref{Qpl}b. Currently, squeezing with microwave fields has reached 10dB~\cite{PRLMicrowaveSqueezing}. Minimising (\ref{sqe}), optimal squeezing is found when $r=\ln({|\Delta_c/\kappa|})/2$. Optimal squeezing at $\phi=\pi/2$ is given by,
 \begin{equation}
  \Delta(\hat X_{\frac{\pi}{2}})_{\textrm{Op}}^2\approx\left|\frac{\kappa}{\Delta_c}\right|\frac{1}{2\left(1+\frac{\kappa^2}{\Delta_c^2}\right)}+\frac{\sigma_p}{2}\left(\frac{\kappa^2}{\Delta_c^2}\right)\left(\Delta(\hat n_+)^2+\Delta(\hat n_-)^2+O(\tilde t_0)\right).
 \end{equation}
Therefore, optimal squeezing is limited strongly by how small is the ratio $|\kappa/\Delta_c|$. Note that each coherent state component is squeezed stronger by a factor of $1/R$, squeezing in each component $\sim 50\%$ more with respect to the total state, see dashed lines in Fig.\ref{Qpl}(b) and (c). Thus the projection to a single component improves optimal squeezing. This projection is even natural, being a consequence of spontaneous symmetry breaking in the system without optical lattice~\cite{EsslingerPRL11}. In the case of $R$ density modes one has in general for $|\kappa/\Delta_c|\leq1$,
\begin{eqnarray}
 \Delta(\hat X_{\frac{\pi}{2}})^2
 &\approx&
\frac{e^{-2r}}{4}+\left(\frac{\kappa^2}{\Delta_c^2}\right)\frac{\sinh(r)\cosh(r)}{1+\frac{\kappa^2}{\Delta_c^2}}
+\frac{\sigma_p}{R}\left(\frac{\kappa^2}{\Delta_c^2}\right)\sum_{q=1}^R\Delta(\hat n_q)^2,
\nonumber
\\
 \end{eqnarray}
 with $\hat n_q$ corresponding number operator of the light-induced density mode component per site.  
In general, projecting to a single component of $R$ modes produces an enhancement factor on squeezing of $1/R$. For example, projecting to a single component with four light induced modes gives a enhancement factor of $75\%$, see Fig.\ref{Qpl}(c) dotted line. Therefore, by incrementing the number of light-induced modes one can optimise squeezing in a single component even though the cavity decay ratio $|\kappa/\Delta_c|$ is not that small.  The general structure of the light-matter properties for arbitrary number of bond and density modes is rather involved as it contains information regarding the correlated phases of matter that emerge. It is instructive to see the effect on the squeezing parameter $r$ for some cases, as we will show in what follows. 

\section{Squeezing parameters and emergent structured phases.}

\subsection{Homogenous light scattering.}

When atoms scatter light homogeneously ($J_{B,\varphi}=J_B$, $J_{D,\varphi}=J_D$, $J_{E,\varphi}=0$), local density imbalance is suppressed. As it has been shown~\cite{MekhovNP2007,LP09,MekhovRev}, in the adiabatic limit SF and MI scatter light differently depending on the properties of the quantum many-body state. As other energy scales become relevant, the additional terms amount to renormalization of the induced interaction in the matter wave coherences, the terms in $\hat B\hat B^\dagger$, such that $\HH_\chi=g_{\mathrm{eff}}\tilde{U}^2\hat B^2/(1-\tilde{U}^2)$ for $|\tilde{U}|<1$. This allows to enhance the effect due to the matter wave coherences via the on-site interaction in the effective Hamiltonian. The squeezing parameter using mean-field approximation is $r=|\langle\hat \xi\rangle|=2 z |g_{\mathrm{eff}}U|J_B^2N_s|\psi^{*2}\langle \hat b_i^2\rangle+\psi^2\langle\hat b_i^{\dagger 2}\rangle-|\psi|^4+\langle \hat n_i^2\rangle- 2 n_i^2-n_i|/\Delta_c^2$, with atom number per site  $n_i=\langle\hat n_i\rangle$ and  the SF order parameter $\psi=\langle \hat b_i\rangle$. Therefore, for a deep classical OL  ($J_B=0$) there will be no squeezing due to $r$. As the classical OL becomes shallower ($J_B\neq0$), the squeezing parameter is maximal for MI and smoothly decreases as we reach the SF state,  see Fig. \ref{transitions}a. This is correlated with the fact that light scattering while illuminating in between density maxima (at the bonds) is maximal in the MI while decreasing as the SF grows~\cite{Wojciech}.

As the number of light induced modes in the matter increases, the induced structures play a substantial role on light squeezing.  For 2 light induced modes, such that their amplitudes alternate sign every other site ($J_{D,\varphi}=\pm J_D$ or $J_{B,\varphi}=\pm J_B$ and $J_{E,\varphi}=2J_D$), we find that the matter induces structure to the squeezing parameter.  As it has been shown~\cite{Santiago} besides from SF, MI the system supports gapped superfluid states, dimer phases,  supersolid (SS) and density waves (DW).

\subsection{Diagonal coupling, illuminating at lattice sites.}

\begin{figure}[tp!]
\begin{center}
\includegraphics[width=0.95\textwidth]{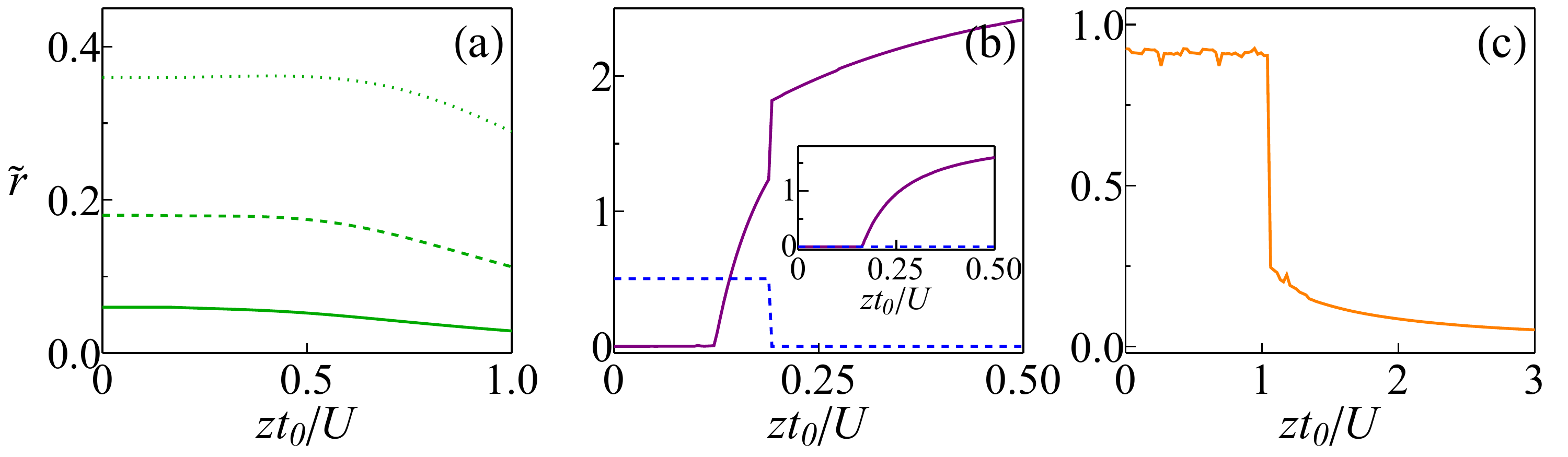}
\end{center}
\caption{ Squeezing parameter across different transitions. (a)  Scattering with a single light induced mode ($J_{D,\varphi}=0,J_{B,0}=J_B, g_{\mathrm{eff}}>0$), as the system goes from MI to SF the squeezing parameter diminishes while the change becomes more visible as density increases $\rho=1.0 (\textrm{solid}), 2.0 (\textrm{dashed}), 3.0 (\textrm{dotted}))$. 
(b) Two component system for maximum light scattering ($J_{D,\pm}=\pm J_D, J_{B,\varphi}=0,  g_{\mathrm{eff}}<0$) at  $\rho=3/2$ filling. The system goes from DW to SS and to SF as $z t_0/U$ increases, dashed line is the difference in density $\Delta\rho=|\rho_+-\rho_-|$ and solid line is the squeezing parameter $r$. 
(b) Inset,  MI to SF transition ($J_{D,\pm}=\pm J_D, J_{B,\varphi}=0,  g_{\mathrm{eff}}<0$) at $\rho=1$. 
(c) Squeezing parameter $r$ for the supersolid dimer (SSD) to SF transition  ($J_{D,\varphi}=0, J_{B,\pm}=\pm J_B,  g_{\mathrm{eff}}<0$) at $\rho=1.0$. For $zt_0/U\lesssim1$ the system is in SSD while  for $zt_0/U\gtrsim1$ is SF.  
Parameters:(a) $g_\mathrm{eff}=U/N_s$, $J_B=0.05$, $\tilde r= r\Delta_c^2/(2 z|g_{\mathrm{eff}}U|J_B^2N_s)$ (b)  $g_\mathrm{eff}=-0.5 U/N_s$, $J_D=1.0$, $\tilde r= r\Delta_c^2/(2 z|g_{\mathrm{eff}}t_0|J_D^2N_s$); (c) $g_\mathrm{eff}= -25U/N_s$, $J_B=0.1$, $\tilde r=2r\Delta_c^2/(z|g_{\mathrm{eff}}U|J_B^2N_s)$. $N_s=100$, $z=6$. }
\label{transitions}
\end{figure}

Without bond ordering ($J_{B,\varphi}=0$, $J_{D,\varphi}=\pm J_D$), the squeezing parameter  is different for SS, DW, SF and MI phases. In mean field theory, the squeezing parameter is proportional to the product of SF order parameters in each light induced component, $r=2 z |g_{\mathrm{eff}}| t_0J_D^2N_s(\psi_+^*\psi^{\phantom{*}}_-+c.c.)/\Delta_c^2$ where $\psi_\pm$ correspond to each light induced mode component.  Thus, for an insulating state (DW or MI) $r=0$ while for a SF $\psi_+=\psi_-$ and in the SS state $\psi_+\neq\psi_-$. Indeed, as the onsite interaction increases i.e. for half integer fillings when atoms scatter light maximally to reach the ground state of $\HH_{\mathrm{eff}}$ ($g_{\mathrm{eff}}<0$), light will be squeezed maximally in SF, while as SS emerges, it will diminish until reaching the DW state where no squeezing is possible, see Fig. \ref{transitions}b. The total coherent state amplitude is $\tilde\alpha_{\pm}\neq0$ when DW order is present, while  $\tilde\alpha_{\pm}=0$ in the SF or MI. Moreover, when atoms scatter light minimally to reach the ground-state ($g_{\mathrm{eff}}>0$), the squeezing parameter is different for SF and gapped SF states. The squeezing parameter for a quantum superposition (QS) state~\cite{Santiago,Atoms} is  $r_{QS}=4 z |g_{\mathrm{eff}}|t_0J_D^2N_s(m+1)(n_i-m)(1+m-n_i)/\Delta_c^2$ for incommensurate fillings  $m<n_i<m+1$ with $m$  positive integer. Thus, for a gapped SF $r\leq r_{QS}$ while for a normal SF state $r> r_{QS}$.

\subsection{Emergent bond order.}

In addition, dynamical terms can induce bond ordering due to the emergent coupling $J_{E,\varphi}$ as $\tilde{U}$ increases. Emergent bond ordering due to density coupling occurs because products of weighted bond and bond current operators modify the effective Hamiltonian via $\HH_\chi$. These terms arise because on-site interaction and tunneling do not commute in general with the light-induced long-range interaction. The new terms that appear in the effective Hamiltonian favour density imbalance as $|\tilde{U}|$ increases and modify the coupling of matter wave coherences with it. Explicitly, we have to order $O(\tilde{t}_0^2\tilde{U}^2)$, 
\begin{equation}
\HH_\chi\approx2g_{\mathrm{eff}}J_D^2\tilde{t}_0^2[(1-\tilde{U}^2)(\hat C_0^\dagger\hat J_0^{\phantom{\dagger}}+h.c.)+\tilde{U}^2(\hat B_1^\dagger\hat B_1^{\phantom{\dagger}}+h.c.)]
\end{equation}
The current operators $\hat J_0=\sum_{\varphi}\hat J_{0,\varphi}$ are structureless but the weighted bond operators $\hat B_{1}=\sum_{\varphi}\hat S_{1,\varphi}=\sum_{\langle i,j\rangle\in\varphi}(\hat n_j-\hat n_i)(\hat b^\dagger_i\hat b^{\phantom{\dagger}}_j+\hat b^\dagger_j\hat b^{\phantom{\dagger}}_i)$ induce a staggered field between bonds as density varies between every other site and the difference between atom populations can alternate sign. Essentially, the density variation acts as an additional dynamical diffraction element that affects the interference of the matter waves in between density maxima. The matter waves in other to compensate the staggered field and optimize the energy in the effective Hamiltonian acquire a phase pattern between adjacent sites.  This translates in the formation of dimer states. The difference in phase of the matter waves $\Delta\phi\neq0$.  Thus, for $g_{\mathrm{eff}}<0$, bond ordering will occur and dimer physics~\cite{Santiago} will emerge even in a deep optical lattice. As a consequence, 4 bond light induced modes will form  leading to a superposition of 4 light-matter correlated squeezed coherent states. The squeezing parameter can be cast as  $r= z |g_{\mathrm{eff}}| t_0J_D^2N_s[\phi_1+\phi_3+(\phi_3+\phi_4)\cos(\Delta\phi)]/\Delta_c^2$ with $\phi_q=|\psi_q^*\psi_{q+1}|$, $\Delta\phi=\arg(\psi_2)-\arg(\psi_{3})=\arg(\psi_4)-\arg(\psi_{1})$ and $\psi_q$ the order parameter of each effective induced mode. Therefore, the squeezing parameter of light inherits the structure due to bond-ordering even in a deep optical lattice. Therefore, the  interplay between short-range processes and the long-range cavity induced interaction leads to the emergence of physics absent in the classical optical lattice and the adiabatic light limit in this configuration. This implies a new alternative for the design of Hamiltonians containing dimer physics in analogy with spin-liquid Hamiltonians~\cite{Balents}.

\subsection{Off-diagonal coupling, illuminating in between lattice sites.}

In the case with only off-diagonal light-matter coupling ($J_{D,\varphi}=0$, $J_{B,\varphi}=\pm J_B$, $J_{E,\varphi}=0$) for maximal light scattering ($g_{\mathrm{eff}}<0$), the squeezing parameter is different between SF, superfluid dimer (SFD), supersolid dimer (SSD) and SS states. The effective interaction strength in the adiabatic limit gets renormalized by the term $\HH_\chi=g_{\mathrm{eff}}\tilde{U}^2(\hat B\hat B^\dagger+\hat B^\dagger\hat B))/2$  to order $O(\tilde{U}^2)$. The squeezing parameter in mean-field approximation and using the typical statistical properties of the states~\cite{PRA2007} can be estimated as $r\approx z |g_{\mathrm{eff}} U|J_B^2N_s|\phi_1^2+\phi_3^2+(\phi_2^2+\phi_3^2)\cos(2\Delta\phi)- n_A n_B|/(2\Delta_c^2)$, where the populations for each dimer are $n_A$ and $n_B$. In contrast to diagonal coupling ($J_{D,\varphi}\neq 0$), the normal SF state  $r$  is minimal ($r\to0$), as $\phi_1=\phi_2=\phi_3=\phi_4=|\psi|^2$, $n_A=n_B= 2 n_0\approx2|\psi|^2$ and $\Delta\phi=0$. In SFD, bond ordering occurs, thus $\phi_1=\phi_3$, $\phi_2=\phi_4$, $n_A=n_B=2 n_0$,  $\Delta\phi\neq0$ with $r\neq0$. Typically, dimer states have $\pi/2<\Delta\phi\leq\pi$, thus $\cos(2\Delta\phi)<0$ depending on the parameters chosen for the system. For SSD, bond ordering and density modulation occurs, then  $\phi_1\neq\phi_3$, $\phi_2\neq\phi_4$, $n_A\neq n_B$, $\Delta\phi\neq0$. Thus,  $r$ is maximal as  $n_A n_B<4n_0$, see Fig. \ref{transitions}c. Therefore, as bond ordering occurs and DW order emerges $r$ is different from zero. For minimal light scattering ($g_{\mathrm{eff}}>0$) one has direct information regarding SS order solely due light-matter quantum correlations. The squeezing parameter has considerably simpler structure with respect to dimer phases since, $r\approx z|g_{\mathrm{eff}} U|J_B^2N_s(n_+-n_-)^2/\Delta_c^2$ with $n_\pm$ the number of atoms in each light induced mode per site. The squeezing parameter tracks directly the emergence of DW order. Therefore, for SS $r\neq0$ and for homogenous SF $r=0$, while the coherent state amplitudes are $\alpha_\pm=0$ but the number of photons is $\langle\hat a^\dagger\hat a\rangle\neq0$.

\section{Effective Master Equation.} 

Beyond the $|\kappa/\Delta_c| \ll 1$ limit, we use the methods of quantum optics~\cite{Walls, Noise}  and we find the effective master equation for the system as
\begin{eqnarray}
\frac{\mathrm{d} \tilde\rho}{\mathrm{d} t}&=&-\frac{i}{\hbar}[\HH_{\mathrm{eff}},\tilde\rho]+\frac{g_{\mathrm{eff}}\kappa}{\Delta_c}\big(2\hat G^\dagger\tilde{\rho}\hat G+[\hat G^\dagger\hat G,\tilde\rho]_+\big)
\\
\tilde\rho&=&\sum_{\varphi_q,\varphi_l}p_{q,l}\beta_{\varphi_q}\beta_{\varphi_l}|\tilde\alpha_{\varphi_q},\xi_{\varphi_q}\rangle_a|\varphi_q\rangle_b \;_b\langle \varphi_l|_a\langle \xi_{\varphi_l}, \tilde\alpha_{\varphi_l} |
\nonumber
\end{eqnarray}
where $\tilde\rho$ is the density matrix, $[\cdot,\cdot]_+$ is the anti-commutator {  and $p_{q,l}$ are the matter coefficients (probabilities) that can describe either a pure or mixed state}. The second term in the master equation is the effective Liouvillian which includes dissipation. Measurement back-action  beyond the $|\kappa/\Delta_c|\ll1$, $|U/\Delta_c|\ll1$ and $|t_0/\Delta_c|\ll1$ limits can be devised by using $\hat G$ as the effective jump operators for quantum trajectories. {   The operator $\hat G$ is related to the coherent state amplitudes of light via $\tilde \alpha_{\varphi_q}|\Psi\rangle=\hat G|\Psi\rangle$. In the above the Markov approximation is implied in the limit when $T=0$~\cite{Walls,Noise}.    }
This allows to consider the effect of measurement back-action, the role of local processes and their interplay due to light-induced non-local interactions simultaneously. This opens a new venue for exploration regarding the design of global structured dissipation channels and measurement induced projection and state design of non-trivial quantum correlated states~\cite{Gabriel,GabrielAFM,WojciechNHQZ}, as well as, control~\cite{BuchleitnerControl,ShersonControl} and the transition to classicality~\cite{Tclassicality}. As it has been shown, this can greatly enhance and optimise the desired quantum properties of light by design.

\section{Conclusions.}

We have shown that quantum optical lattices offer a new tool to engineer a generalised class of states that are a non trivial superposition of structured squeezed coherent states of light entangled with matter. These states are entangled with the matter at the fundamental level due to the structure of quantum many-body matter states.  We have demonstrated that breaking symmetries by design one can induce structure to the parameters that control the nonclassical features of light. This has been shown to be accessible via quantities such as, the photon number and the quadratures of light.  We have shown how the quantum properties of light contain the information of matter-field coherences, density patterns of matter and light-matter quantum correlations. Thus the properties of strongly correlated phases of matter get imprinted on the quantum properties of  light. Moreover, we have found that the interplay between induced long-range processes and ordinary short-range atomic processes lead to the modification of the effective Hamiltonian of the system. We have obtained that the effect of local processes can be used to generate delocalized dimer phases due to the dynamical properties of light even in deep optical lattices. Moreover, one can optimise these nonclassical features  depending on cavity parameters and the structure imprinted to the matter that gets transferred to the light-induced mode structure. Additionally, the non-trivial light-matter correlated states that arise can be used to design dissipation channels via the effective master equation and measurement back-action.  A pathway to study the behaviour we describe is to combine several recent experimental breakthroughs:  detection of light scattered from ultracold atoms in OL was performed, but without a cavity~\cite{Weitenberg2011,KetterlePRL2011} and  BEC was trapped in a cavity, but without a lattice~\cite{EsslingerNat2010,HemmerichScience2012,ZimmermannPRL2014}. As our treatment of the system is based on off-resonant scattering, this is not sensitive to a detailed atomic level structure. Therefore, our treatment applies to analogous arrays of natural or artificial quantum objects such as: spins, fermions, molecules (including biological ones)~\cite{LPhys2013}, ions~\cite{Ions2012}, atoms in multiple cavities~\cite{ArrayPolaritons2006}, semiconductor~\cite{SQubits2007} or superconducting qubits~\cite{JCQubits2009}.

\section*{Aknowledgements}
The work was supported by the EPSRC (EP/I004394/1).

  \end{document}